\documentclass[letter]{aa} % for the letters 
\usepackage{graphicx}
\usepackage{longtable}
\usepackage{txfonts}
\begin{document}
%
%   \title{GRB\,050408: \\Shaping the multiband light curve of an X-ray rich burst
%   \title{X-ray rich GRB\,050408: The bumpy afterglow of an off-axis burst
   \title{Extensive multiband study of the X-ray rich GRB\,050408
   \thanks{Based on observations collected at SAO, La Silla, Roque de los Muchachos, 
Haleakala, Kitt Peak, Cerro Tololo, T\"UBITAK, Kiso, Observatorio de Sierra Nevada, 
Plateau du Bure, GMRT and RATAN-600.}}
   \subtitle{A likely off-axis event with an intense energy injection}

   \author{
        A.~de Ugarte Postigo \inst{1}
   \and T.A.~Fatkhullin \inst{2}
   \and G.~J\'ohannesson \inst{3}
   \and J.~Gorosabel \inst{1}
   \and V.V.~Sokolov \inst{2}
   \and A.J.~Castro-Tirado \inst{1}
   \and Yu.Yu.~Balega \inst{2}
   \and O.I.~Spiridonova \inst{2}
   \and M.~Jel\'{\i}nek \inst{1}
   \and S.~Guziy \inst{1,4}
   \and D.~P\'erez-Ram\' irez \inst{5}
   \and J.~Hjorth \inst{6}
   \and P.~Laursen \inst{6}
   \and D.~Bersier \inst{7}
   \and S.B.~Pandey \inst{1,8}
   \and M.~Bremer \inst{9}
   \and A.~Monfardini \inst{7}
   \and K.Y.~Huang \inst{10}
   \and Y.~Urata \inst{11,12}
   \and W.H.~Ip \inst{10}
   \and T.~Tamagawa \inst{11}
   \and D.~Kinoshita \inst{12}
   \and T.~Mizuno \inst{13}
   \and Y.~Arai \inst{13}
   \and H.~Yamagishi \inst{13}
   \and T.~Soyano \inst{14}
   \and F.~Usui \inst{15}
   \and M.~Tashiro \inst{16}
   \and K.~Abe \inst{16}
   \and K.~Onda \inst{16}
   \and Z.~Aslan \inst{17,18}
   \and I.~Khamitov \inst{17}
   \and T.~Ozisik \inst{17}
   \and U.~Kiziloglu \inst{19}
   \and I.~Bikmaev \inst{20,21}
   \and N.~Sakhibullin \inst{20,21}
   \and R.~Burenin \inst{22}
   \and M.~Pavlinsky \inst{22}
   \and R.~Sunyaev \inst{22}
   \and D.~Bhattacharya \inst{23}
   \and A.P.~Kamble \inst{23}
   \and C.H.~Ishwara Chandra \inst{24}
   \and S.A.~Trushkin \inst{2}
    }
   \offprints{A. de Ugarte Postigo, deugarte@iaa.es}

   \institute{
           Instituto de Astrof\'{\i}sica de Andaluc\'{\i}a (IAA-CSIC),
           Apartado de Correos 3004, E-18080 Granada, Spain.{}
           \and
           Special Astrophysical Observatory, Nizhnij Arkhyz, Zelenchokskaya, Karachaevo-Cherkesia, 369167 Russia.{}
           \and
           Science Institute, University of Iceland, Dunhaga 3, IS-107 Reykjav\'{\i}k, Iceland.{}
           \and
           Nikolaev State University, Nikolska 24, 54030, Nikolaev, Ukraine.{}
           \and
           Dpto. de F\'{\i}sica (EPS), Universidad de Ja\'en, E-23071, Ja\'en, Spain.{}
           \and
           Dark Cosmology Centre, Niels Bohr Institute, University of Copenhagen,
           Juliane Maries Vej 30, DK-2100 Copenhagen, Denmark. {}
           \and
           Astrophysics Research Inst., Liverpool John Moores Univ., Twelve Quays House, Egerton Wharf, Birkenhead, CH41 1LD, UK.{}
           \and
           The UCL Mullard Space Science Laboratory, Holmbury St. Mary, Dorking, Surrey, RH5 6NT, UK. {}
           \and
           Institut de Radio Astronomie Millim\'etrique (IRAM), 300 rue de la Piscine, 38406 Saint-Martin d'H\`eres, France. {}
           \and
           Institute of Astronomy, National Central University, Chung-Li 32054, Taiwan, Republic of China.{}
           \and
           Institute for Physics and Chemical Research (RIKEN), Wako, Saitama 351-0198, Japan.{}
           \and
           Tokyo Institute of Technology, Ookayama, Meguro, Tokyo 152-8550, Japan.{}
           \and
           Department of Astronomy and Earth Sciences, Tokyo Gakugei University, Koganei, Tokyo 184, Japan.{}
           \and
           Kiso Observatory, Institute of Astronomy, University of Tokyo, Mitake-mura, Kiso-gun, Nagano 397-0101, Japan.{}
           \and
           Japan Aerospace Exploration Agency, Institute of Space and Astronautical Science, Sagamihara, Kanagawa 229-8510, Japan.{}
           \and
           Saitama University, Sakura-ku, Saitama 338-8570, Japan.{}
           \and
           T\"UBITAK National Observatory, Akdeniz \"Universitesi, 07058, Antalya, Turkey.{}
           \and
           Akdeniz University, Physics Department, 07058 Antalya, Turkey.{}
           \and
           Middle East Technical University, Physics Department, Inonu Bulvari, Ankara, 06531, Turkey.{}
           \and
           Departments of Astronomy, Kazan State University, Kremlevskaya Str., 18, Kazan, 420008, Russia.{}
           \and
           Academy of Sciences of Tatarstan, Bauman Str., 20, Kazan, 420111, Russia.{}
           \and
           Space Research Institute (IKI), 84/32 Profsoyuznaya, Moscow, 117997, Russia.{}
	   \and
	   Raman Research Institute, Bangalore 560 080, India.{}
	   \and
           National Centre for Radio Astrophysics, Ganeshkhind, Pune, 411007, India.{}
           \\
%             \email{deugarte@iaa.es}
%             \thanks{}
             }

   \date{}

% \abstract{}{}{}{}{} 
% 5 {} token are mandatory
 
  \abstract
   {}
   {Understand the shape and implications of the multiband light curve of 
GRB\,050408, an X-ray rich (XRR) burst.}
   {We present a multiband optical light curve, covering the time from the onset 
of the $\gamma$-ray event to several months after, when we only detect the host 
galaxy. Together with X-ray, millimetre and radio observations we compile what, 
to our knowledge, is the most complete multiband coverage of an XRR burst afterglow 
to date.}
   {The optical and X-ray light curve is characterised by an early flattening 
and an intense bump peaking around 6 days after the burst onset. We explain the 
former by an off-axis viewed jet, in agreement with the predictions made for XRR 
by some models, and the latter with an energy injection equivalent in intensity 
to the initial shock. The analysis of the spectral flux distribution reveals an 
extinction compatible with a low chemical enrichment surrounding the burst. 
Together with the detection of an underlying starburst host galaxy we can 
strengthen the link between XRR and classical long-duration bursts.}
   {}

   \maketitle
%
%________________________________________________________________

\section{Introduction}

X-ray flashes (XRFs) where first identified by {\it
Beppo-SAX} (\cite{hei01}) as those bursts detected by the X-ray camera
but not the $\gamma$-ray monitor. Later studies based on the larger sample
gathered by {\it HETE-2} (\cite{sak05b}) gave a more general (and instrument-
independent) classification and confirmed the intermediate group of
events, the X-ray rich (XRR) class, previously detected by {\it Ginga} (\cite{yos94}) and {\it Granat}/WATCH
(\cite{cas94}). It is now known that long-duration GRBs (LGRBs),
XRRs and XRFs share the same isotropic distribution in the sky, the same
duration range and similar spectrum, with the main difference of having
respectively lower observed spectral peak energy $E_{peak}^{obs}$ in the
$\nu F_{\nu}$ spectrum. They seem to form a continuum and thereby, most
of the proposed models have tried to explain them as a unified phenomena
(see a summary of the different models in \cite{gra05}).

GRB\,050408 was detected by WXM, SXC, and FREGATE aboard {\it HETE-2}
(\cite{sak05}) at 2005 April 08 16:22:50.93 UT (t$_0$ hereafter). With 
an observed peak energy of $\sim$20 keV it was classified as an XRR event. 
The 1.0m Zeiss and 6.0m BTA telescopes at the Special Astrophysical 
Observatory (SAO) in Russia pointed at the position delivered by {\it 
HETE-2} through the GCN (GRB Circular Network) and detected the optical 
afterglow (\cite{deu05a}) coincident with the possition of the X-ray 
afterglow detected by {\it Swift}/XRT, which began observing 42 minutes 
after the burst (\cite{wel05}). The precise localisation allowed further 
optical (Covino et al. \cite{cov07}) and spectroscopic (\cite{ber05,pro05}) 
observations, this latter ones detemined a redshift of $z$=1.236.

In Sect. 2 we present the  observations and the reduction methods that
have been used for the analysis of the data. Sect. 3 describes
the results that   have   been   obtained, including observations of the
host galaxy and modelling of the light curve.  Sect. 4 discusses the
implications of the  analysis of the light curve.

%__________________________________________________________________

\section{Observations and data reduction}

For this work we have compiled over 60 photometric measurements in {\it U, B,
V, Rc \rm and \it Ic} bands from 12 telescopes. The images where reduced using
standard techniques based on IRAF \footnote{IRAF is distributed by the National
Optical Astronomy Observatories, which is operated by the Association of
Universities for Research in Astronomy, Inc. (AURA) under cooperative agreement
with the National Science Foundation} and JIBARO (\cite{deu05b}).

The burst happened during night time in Japan, where a fast follow up was 
carried out. The very wide field camera, WIDGET was 
monitoring the field of view of {\it HETE-2} when the event was reported but 
detected no optical emission before, during or after the gamma-ray emission down to an 
unfiltered limiting magnitude of 9.7 (all limits given throughout the paper 
are 3-$\sigma$). The 1.05m KISO Schmidt telescope pointed to the error box 20
minutes after the burst but failed to detect the afterglow. Finally, the 1m LOT 
telescope observed the field 55 minutes 
after the burst, images that later served to confirm the afterglow (\cite{hua05b}).
The discovery of the afterglow was made with the data of the 1.0m Zeiss and
6.0m BTA telescopes in Russia (\cite{deu05a}), starting 115 minutes after
the burst, when observations became possible from that site.

Further observations were performed from the 1.5m Russian-Turkish telescope, in 
T\"UBITAK National Observatory, the 4.0m Blanco telescope in Cerro Tololo, the 
4.0m Mayall telescope in Kitt Peak, the 2.0m Faulkes Telescope North (FTN) in 
Haleakala and the 3.5m Telescopio Nazionale Galileo (TNG) in la Palma. A specially 
intense multiband campaign was carried out from the 1.54m Danish telescope in La 
Silla, where daily observations were obtained during the first 8 days following 
the burst onset.

Finally, 8 months after the burst, deep observations were made from the 
3.5m telescope at Calar Alto. In these images we detect the host galaxy of the 
burst in {\it B} and {\it Rc} bands and impose a limit in {\it Ic} band.

Optical photometric calibration is based on the observation of several standard 
fields (\cite{lan92}) using the 1.54m Danish telescope at La Silla and the 1.5m 
telescope at Sierra Nevada Observatory. From these observations we derive 12 
secondary standards of different brightnesses. A log with the observations and 
the calibration stars are given as online material.

Our dataset is completed with several millimetre and radio limits. 6 epochs of 
millimetre observations were carried out with the 6-antenna Plateau de Bure 
interferometer (PdB, \cite{gui92}). 
No detection was obtained in either of the 1 mm or 3 mm bands, although a 
3-$\sigma$ signal was found on the phase center in both observing bands 
on April 18. Careful re-analysis of the data did not reveal these signals 
as instrumental artifacts. Based on the extreme spectral slope
and the non-detection on April 19, we conclude that this result is either 
due to a statistical fluctuation or an unusual event of interstellar 
scattering at high galactic latitude, and not due to a source-intrinsic variation.
Data calibration was done using the GILDAS software 
package\footnote{GILDAS is the software package distributed by the IRAM 
Grenoble GILDAS group.} using MWC349 as primary flux calibrator and 3C273 as 
amplitude and phase calibrator. Further observations where obtained 13 days 
after the burst  at 1.28 GHz from GMRT and at 8.4 GHz from RATAN-600. 

%______________________________________________ 

\begin{figure}[ht!]
\centering
\includegraphics[width=\hsize]{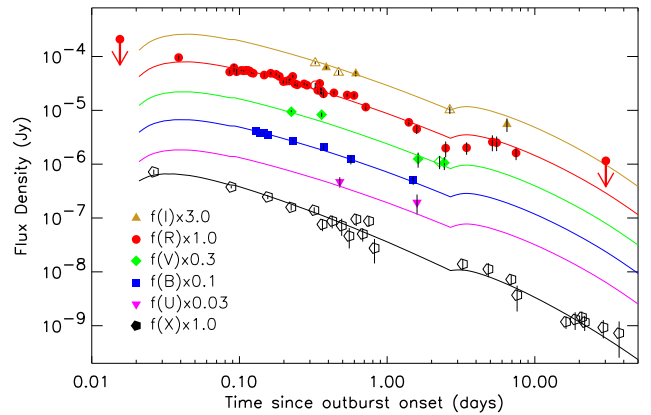}
\protect\caption[ ]{Multiwavelength light curve of the GRB\,050408
afterglow in the observer frame. The lines show the best fit of a fireball model with one energy
injection (at 2.9 days) seen off-axis (see text for details). Our observations are
plotted with filled symbols, while the ones obtained from the literature
are represented by empty ones, this convention is used for all the figures.\label{lc}}
\end{figure}

%\vspace{-1.0cm}

\section{Results}

\subsection{Light curve}

In order to put together all the radio, millimetre, near infrared (nIR), optical and
X-ray data available (including data from Foley et al. \cite{fol06} and
\cite{sod05a,sod05b}), we have determined the corresponding flux density
values for all observations. X-ray afterglow counts, obtained from Foley
et al. (\cite{fol06}) have been converted and corrected for hydrogen
column extinction using WEBPIMMS\footnote{http://heasarc.nasa.gov/Tools/w3pimms.html} 
taking as spectral model a powerlaw with a slope of $\beta_X$=1.16 and a 
column density of N$_{\rm H}$=$0.25\times10^{22} cm^{-1}$ (\cite{nou06}, Chincarini et al. \cite{chi05}).
The optical data have been corrected for galactic reddening (using an E(B-V)=0.026, Schlegel et al.
\cite{Schl98}) and intrinsic extinction (see Sect. 3.2). The
measured/estimated host galaxy flux has been subtracted
from the data to obtain the clean afterglow
flux (see Sect. 3.3). The conversion of the optical data to flux density was done
using  the transformations given by Fukugita  et al.  (\cite{Fuku95})
for the optical and by Allen et al. (\cite{Allen00}) for the nIR. The
resulting light curves are shown in Fig. 1. Note the intense bump, 
rising at $\sim$3 days and peaking at $\sim$6 days, both in optical 
and X-rays. These kind of fluctuations have already been detected in the
light curves of LGRBs and short-duration bursts (SGRBs) (\cite{deu05c,deu06}).

\subsection{Study of the optical-nIR SFD}

We  have constructed  the $UBVRcIcJHK$-band  Spectral  Flux distribution
(SFD)  of  the   afterglow  0.6  days  after  the   burst,  when  near
simultaneous  optical  and  nIR  observations  were  available.   The
$UBVRcIc$-band  magnitudes from this work were complemented with the 
$JHK$-band values reported by Foley et al. (\cite{fol06}). Synchronisation
to a  unique timing is done by  assuming a powerlaw with an index
of $\alpha$ = 0.7 (F $\propto t^{-\alpha}$), as derived from a linear
fit of the nearby multiband data of the afterglow.

The fluxes are  used for fitting an extincted  powerlaw ($F_{\nu} \propto
10^{-0.4A_{\nu}}\nu^{-\beta}$) with 3  different 
extinction laws: Milky  Way (MW), Large Magellanic Cloud  (LMC) and Small 
Magellanic Cloud (SMC) as described by Pei (\cite{Pei92}).  This allows us  
to obtain A$_{\rm V}$ and $\beta$  simultaneously.  The results of these 3 
fits are complemented with an unextincted powerlaw case (NE), see Fig. 2 
and Table \ref{sedtable}. The best fit to the SFD of the afterglow is 
obtained when considering a SMC extinction law ($\chi^{2}/d.o.f.$ = 5.0/5). 
This is consistent with what has been previously found for other
LGRB afterglows (\cite{kan06}).

%\vspace{-0.5cm}

\begin{table}[h]
\begin{center}
\caption{Results of the SFD fitting at 0.6 days for different
extinction laws. \label{sedtable}} \scriptsize{
\begin{tabular}{c c c c}
\noalign{\smallskip} \hline\hline \noalign{\smallskip}
 Extinction Law &  $\beta$       &     $A_{\rm V}$    & $\chi^{2}/d.o.f.$ \\
\noalign{\smallskip} \hline \noalign{\smallskip}
 MW             &  1.85$\pm$0.30 & -0.18$\pm$0.22 &    24.5/5\\% 4.9      \\
 LMC            & -0.12$\pm$0.48 &  1.19$\pm$0.32 &    11.5/5\\% 2.3      \\
 SMC            &  0.28$\pm$0.33 &  0.73$\pm$0.18 &     5.0/5\\% 1.0      \\
 NE             &  1.62$\pm$0.07 &        0       &    21.0/5\\% 4.2      \\
\noalign{
\smallskip} \hline \hline\end{tabular}
} \normalsize \rm
\end{center}
\end{table}

%\vspace{-1.0cm}

\begin{figure}[ht!]
\centering
\includegraphics[width=\hsize]{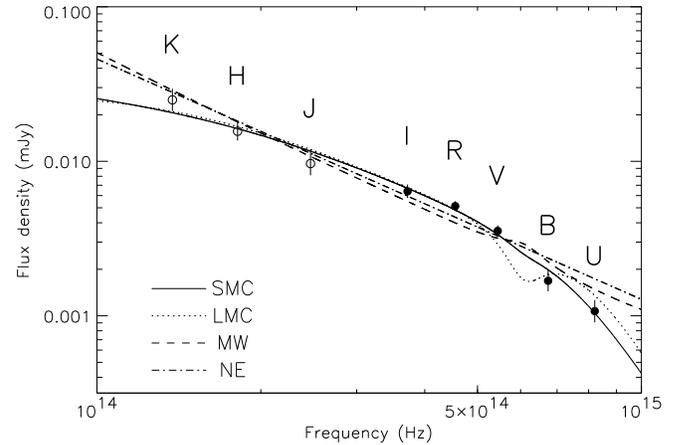}
\protect\caption[SFD]{Spectral flux distribution of the afterglow 0.6 days after the burst onset in the observer frame.
The different lines represent results from fitting the data to various extinction laws: Small
Magellanic Cloud (SMC), Large Magellanic Cloud (LMC), Milky Way (MW) and No Extinction (NE).
\label{lc}}
\end{figure}

%\vspace{-1.0cm}

\subsection{The host galaxy}
Several months after the gamma-ray event we revisited the GRB field
with the 3.5m telescope at Calar Alto Observatory in order to search for the host
galaxy. Images were obtained in {\it B}, {\it Rc} and {\it Ic} bands, yielding a faint
detection in {\it B} and {\it Rc} and imposing a limit on {\it Ic}. We derive galaxy 
colour indices of $(B-Rc) = 0.7 \pm 0.5$ and $(Rc-Ic) \le 0.73$. These values are 
corrected for Galactic extinction. We have compared these values to the ones derived 
from the templates computed by Kinney et al. (\cite{kin96}) for
a wide variety of galaxy types. We may conclude that only starburst galaxy 
templates are consistent with them. The best correlation is obtained
with the starbust 2 template, with an intrinsic extinction of E(B-V) = 0.16.

\subsection{Modelling of the multiband data}

Using the model and methods described by \cite{joh06} we fitted the multiband
observations of the afterglow (galaxy subtracted) to a fireball model with
energy injections, viewed both on-axis and off-axis (with varying
viewing angles). At least one injection is needed in order to account for the
bump seen at 6 days which would carry as much energy as the initial shock.
Another characteristic of the light curve is a flattening of the early
light curve, seen in Rc and X-rays during the first hours of the burst,
which has already been reported by Foley et al. (\cite{fol06}). This can
be explained either by an early energy injection (single or continous), or an outflow with a low initial Lorentz 
factor, or as the result of an off-axis viewed burst. Similar early behaviour has
already been found in other bursts (\cite{nou06,zha06}).

Our preferred scenario (giving the best fit) describes the burst as a collimated 
($\theta_0=2.7^\circ$) fireball seen off axis ($\theta_v=1.45 \theta_0$) expanding 
into a uniform low density environment ($n_0=0.01$ cm$^{-3}$) with an electron index 
$p=2.03$ and having an additional energy injection after 2.9 days with 1.2 times 
the initial energy. No further injections are needed to explain the light curve 
with the available amount of data. From the fit we obtain a $\chi^2/d.o.f. = 158.1/93$. 
The jet break, as defined by Sari et al. (\cite{sar99}), would be 
expected initially at 1.6 days, or at 3.8 days due to the energy injection. However, 
due to the effect of the equal arrival time surface, it is further delayed to 
approximatelly 30 days. Fig. 3 shows the radio to X-ray SFD predicted by our model 
for 3 epochs, together with observational data, host subtraced and corrected for 
galactic and intrinsic extinction.

%\vspace{-0.5cm}

\begin{figure}[ht!]
\centering
\includegraphics[width=\hsize]{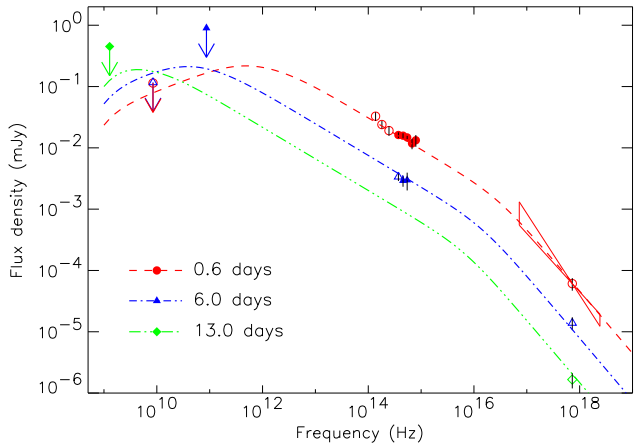}
\protect\caption[SFD]{Spectral Flux Distribution of the afterglow from
radio to X-rays 0.6, 6.0 and 13.0 days after the burst in the observer frame. Several 3-$\sigma$ upper limits from 
radio and millimetre observations are plotted.
\label{sed2}}
\end{figure}

%\vspace{-0.5cm}

%__________________________________________________________________

\section{Discussion}

The optical-nIR SFD shows a clear curvature, implying a need for extinction
along  the  afterglow line  of  sight in  the  host  galaxy. The  only
reliable fit ($\chi / d.o.f.= 5.0/5$)  is based on a SMC extinction law.
This result points towards  a low  stage of  chemical enrichment in the 
region of the progenitor, as is usually found for LGRBs (\cite{kan06}). 
The detection of a starburst host galaxy is also a common feature with 
most LGRBs (\cite{fru06}), facts that once again favour the hypothesis 
of shared nature between XRR bursts and LGRBs.

The optical spectral index obtained from this fit ($\beta_o =0.28\pm0.33$) 
and the X-ray one ($\beta_{X}  = 1.14\pm0.19$, \cite{nou06})
are consistent with a synchrotron spectra in which the cooling break
frequency ($\nu_c$) is located between optical and X-rays. A linear fit
of the optical ($Rc$) and X-ray data between 0.1 and 1.0 days (where
there is more data available and the light curve seems more stable) returns 
temporal slope values of $\alpha_{o} = 0.69 \pm 0.04$ and $\alpha_X = 0.99 
\pm 0.21$. These numbers, together with the optical and X-ray spectral slopes, 
are consistent with a standard fireball model (Sari et al. \cite{sar99}) in which a 
relativistic outflow is expanding in a uniform density environment in 
the slow cooling regime with an electron power law distribution index of 
$p \sim 2.0$.

A more complex multirange model, confirms these results and is used to account 
for the bump that has been detected to peak at $\sim$ 6 days by allowing for refreshed 
shocks. This fluctuation is simultaneously observed in optical and X-rays and can 
be explained by an energy injection of the order of the initial shock. 
This achromaticity and the simultaneity at both sides of $\nu_c$ rules out 
other explanations such as a density fluctuation, a dust echo or a supernova 
bump (which could also be ruled out by amplitude and onset time). Other 
explanations involving an refreshed energy release such as a double jet (\cite{ber03}) or a 
patchy shell (\cite{mes98}) can not be discarded. This injection delays 
the break, that would be expected for about 1.6 days, to about 30 days and, 
mainly due to the effect of the equal arrival time surface, transforms it 
to a very smooth break that expands over a decade in time.

To explain the flattening seen in the earliest points of the light curve
we have studied the case of a collimated fireball seen off-axis, as
predicted by some unified models (\cite{yam02}) that simultaneously intend to explain 
LGRBs, XRR bursts and XRFs by only varying the viewing angle. Our fit accounts 
reasonably well for the multiband and long scale behaviour of the light curve. 
However, the fits obtained with an on-axis model with an additional early 
injection or a low initial gamma factor (dirty fireball) can also interpret the data (although returning 
worse fits) and can not be ruled out.

Regarding the energetics of the afterglow we find that, with an observed peak energy of
$\sim$20 keV and a fluence of $\sim~3.3\times10^{-6}$erg cm$^{-2}$ (2-400 keV)
it has an isotropic equivalent energy release in $\gamma$-rays $E_{\gamma,iso}\gtrsim 
1.3\times10^{52}$ erg, at least 6 times greater than the predicted by E$_{peak}$-E$_{iso}$ 
relation (\cite{ama06}).

%__________________________________________________________________

%\section{Conclusions}

%   \begin{enumerate}
%\item GRB\,050408 is, to our knowledge, the first X-ray rich burst for which an extensive photometric follow
%up has been carried out in optical and X-rays. Additional datapoints in NIR and some
%limits in radio and milimetric wavelengths complete the dataset.

%\item It is a bright burst for which $E_{\gamma,iso}$ is at least 6 
%times greater than predicted by the Amati relation for its E$_{peak}$.

%\item A SFD study shows that the afterglow spectra can be modelled by a powerlaw 
%with a slope of 0.28$\pm$0.33 and a SMC extinction of $A_{\rm V}=0.73\pm0.18$
%favouring a low metallicity in the surroundings of the progenitor. Late time 
%observations show an underlying starburst host galaxy. Both SMC extinction
%and a starburst host galaxy are typically found in LGRBs, favouring unification models.

%\item We have also performed a complete modelling of the afterglow in the context
%of a collimated fireball model. This model includes an energy injection of the 
%order of the initial shock energy, explaining the achromatic bump peaking at ~ 6 days. 
%The flattening during the first hours of the burst, reported by Foley et al. 
%(\cite{fol06}) is explained through the off-axis view of the jet. This is expected 
%from some theories that unify the nature of LGRBs, XRFs and XRR bursts by only
%changing the viewing angle (\cite{yam02}).
% \end{enumerate}

We encourage polarimetric observations of XRR bursts and XRF events (i.e. \cite{gor06}) 
as they will be extremely useful to better understand the physics and geometry of the 
emission and to discriminate between energy models when explaining the fluctuations 
seen in the light curves.

\begin{acknowledgements}
We acknowledge  the generous  allocation of observing  time by different Time 
Allocation Committees. This work was partially supported by the Spanish MCyT 
under programmes AYA2004-01515 and ESP2005-07714-C03-03 (including FEDER 
funds), RFBR grants 04-02-16300 and 05-02-17744 and grant NSh-784.2006.2.
The Dark Cosmology Centre is supported by the Danish National Research 
Foundation. AdUP acknowledges support from FPU grant AP2002-0446 from the 
Spanish MCyT. GJ acknowledges support from the Icelandic Research Council.
\end{acknowledgements}

%\vspace{-0.5cm}

\Online

\begin{appendix} %First online appendix
\section{Calibration stars}
Optical photometric calibration is based on the observation of several
standard fields (\cite{lan92}) using the 1.54m Danish at La Silla (Chile)
and the 1.5m telescope at Sierra Nevada Observatory (Spain). From these
observations we derive 12 secondary standards of different
brightnesses in the field of the GRB. Identification of these stars is
shown on Fig. A.1, while their photometric values are displayed on Table A.1.

\begin{figure*}
\centering
\includegraphics[]{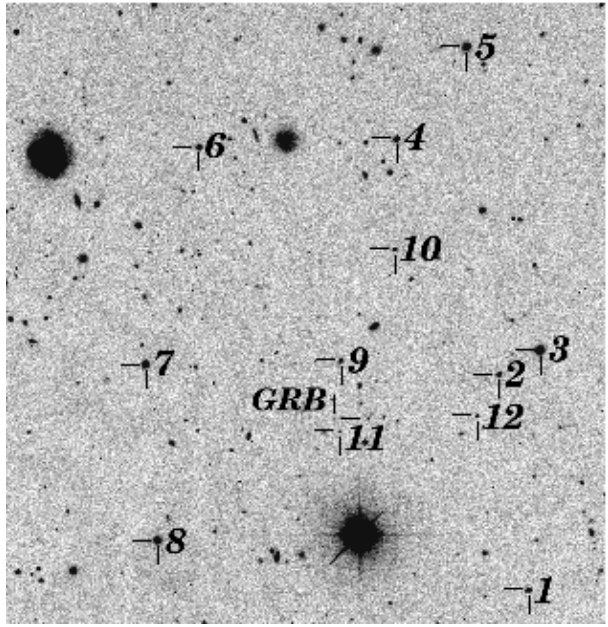}
   \caption{Identification chart of GRB\,050408. The calibration
stars and the afterglow location have been marked. The field of
view is 6.0$'$ $\times$ 6.5$'$, North is to the top and East to
the left.}
      \label{id}
\end{figure*}

\begin{table*}[h]
\begin{center}
\caption{Calibration stars in the field of GRB\,050408, as  marked in Fig. A.1.
\label{calibtab}} \scriptsize{
\begin{tabular}{r c c c c c c c}
\noalign{\smallskip} \hline\hline \noalign{\smallskip}
\#&  RA (J2000) & Dec (J2000) &         U        &         B        &         V        &         R        &         I
 \\
\noalign{\smallskip} \hline \noalign{\smallskip}
1 & 12:02:09.16 & +10:49:24.1 & 19.114$\pm$0.053 & 19.174$\pm$0.024 & 18.601$\pm$0.009 & 18.213$\pm$0.008 & 17.835$\pm$0.024
 \\
2 & 12:02:10.38 & +10:51:37.3 & 18.982$\pm$0.053 & 19.241$\pm$0.024 & 18.776$\pm$0.009 & 18.438$\pm$0.008 & 18.062$\pm$0.025
 \\
3 & 12:02:08.70 & +10:51:52.4 & 19.743$\pm$0.054 & 18.759$\pm$0.023 & 17.167$\pm$0.008 & 16.125$\pm$0.007 & 14.776$\pm$0.022
 \\
4 & 12:02:14.68 & +10:54.02.8 & 18.200$\pm$0.051 & 18.508$\pm$0.023 & 18.028$\pm$0.008 & 17.671$\pm$0.007 & 17.292$\pm$0.023
 \\
5 & 12:02:11.76 & +10:54:59.6 & 18.204$\pm$0.051 & 17.855$\pm$0.023 & 17.089$\pm$0.008 & 16.627$\pm$0.007 & 16.218$\pm$0.022
 \\
6 & 12:02:22.95 & +10:53:57.8 & 18.896$\pm$0.051 & 19.098$\pm$0.023 & 18.638$\pm$0.009 & 18.300$\pm$0.008 & 17.980$\pm$0.024
 \\
7 & 12:02:25.20 & +10:51:43.6 & 20.034$\pm$0.058 & 19.143$\pm$0.023 & 17.681$\pm$0.008 & 16.742$\pm$0.007 & 15.695$\pm$0.022
 \\
8 & 12:02:24.70 & +10:49:54.7 & 17.237$\pm$0.051 & 17.406$\pm$0.023 & 17.020$\pm$0.008 & 16.652$\pm$0.007 & 16.324$\pm$0.022
 \\
9 & 12:02:17.02 & +10:51:45.5 & 20.510$\pm$0.064 & 20.805$\pm$0.030 & 20.216$\pm$0.018 & 19.673$\pm$0.020 & 18.951$\pm$0.041
 \\
10& 12:02:14.76 & +10:52:54.6 &        ---       & 21.885$\pm$0.058 & 20.528$\pm$0.019 & 19.617$\pm$0.014 & 18.631$\pm$0.028 \\
11& 12:02:17.10 & +10:51:02.5 &        ---       & 23.674$\pm$0.051 & 22.376$\pm$0.098 & 21.823$\pm$0.092 & 20.947$\pm$0.137 \\
12& 12:02:11.32 & +10:51:11.3 & 21.129$\pm$0.089 & 21.249$\pm$0.034 & 20.810$\pm$0.023 & 20.507$\pm$0.029 & 20.025$\pm$0.066 \\
\noalign{
\smallskip} \hline \hline\end{tabular}
} \normalsize \rm
\end{center}
\end{table*}

\section{Observations}

In the following tables we use the {\it HETE-2} onset time t$_0$ = 2005 April 08 16:22:50.93 UT.

\begin{table}[h]
\begin{minipage}[t]{\columnwidth}
\begin{center}
\caption{Observations in millimetre wavelengths from the Plateau du Bure interferometer 
(compact 6-antenna configuration on all dates). The errors are based on 
point-source fits in the UV plane to the phase center. \label{mmtable}} \scriptsize{
\renewcommand{\footnoterule}{}
\begin{tabular}{c c c c }
\noalign{\smallskip} \hline\hline \noalign{\smallskip}
(t-t$_0$) (days)&Band (GHz)& Flux (mJ)   & 3-$\sigma$ limit (mJy)\\
\noalign{\smallskip} \hline \noalign{\smallskip}
 3.23           &  86.789  & 0.0$\pm$0.3 & 0.9          \\
 3.23           & 229.068  & 1.0$\pm$1.6 & 4.8          \\
 5.19           & 115.477  & 0.5$\pm$0.8 & 2.5          \\
 5.19           & 232.295  & 3.3$\pm$1.8 & 5.4          \\
 10.34	        & 86.251   & 0.9$\pm$0.3 & 0.9\footnote{The faint detections found on 
this epoch are considered to be due to a statistical fluctuation or to an unusual event
of interstellar scattering at high galactic latitud, taking into account the 
non detection the next night and the extreme spectral slope} \\
 10.34 	        & 232.171  & 8.4$\pm$2.3 & 6.9$^a$\\
 11.19	        & 108.995  & -1.7$\pm$1.7& 5.2\\
 11.19         	& 228.534  & -9.9$\pm$6.5& 19.5\\
 12.40          & 108.995  & -0.9$\pm$0.7& 2.2           \\
 12.40          & 228.534  & 3.6$\pm$2.7 & 8.2           \\
 14.29          & 111.619  & -0.8$\pm$0.4& 1.3           \\
 14.29          & 224.680  & 1.4$\pm$2.1 & 6.3           \\
\noalign{
\smallskip} \hline \hline\end{tabular}
} \normalsize \rm
\end{center}
\end{minipage}
\end{table}

\begin{table}[h]
\begin{center}
\caption{Observations in radio wavelengths from the Giant Metrewave Radio 
Telescope (GMRT) and Radio Astronomical Telescope Academy Nauk (RATAN-600). 
\label{mmtable}} \scriptsize{
\begin{tabular}{c c c c}
\noalign{\smallskip} \hline\hline \noalign{\smallskip}
(t-t$_0$) (days)& Telescope & Band (GHz)& 3-$\sigma$ limit (mJy)\\
\noalign{\smallskip} \hline \noalign{\smallskip}
 13.0   & GMRT & 1.28 & 0.45 \\
 13.0   & RATAN-600 &  8.4 & 5.0 \\
\noalign{
\smallskip} \hline \hline\end{tabular}
} \normalsize \rm
\end{center}
\end{table}

\onllongtab{3}{
\begin{longtable}{cccccc}
\caption{Optical observations carried out for GRB\,050408. The magnitudes are
in the Vega system and are not corrected for Galactic reddening.}\\
\hline
\hline
(t-t$_0$) (days)& Tel. + Inst.   &  Filter  & Texp (s)  &  Mag  & ErMag\\
\hline\hline
\endfirsthead
\caption{Continued.}\\
\hline\hline
(t-t$_0$) (days)& Tel. + Inst.   &  Filter  & Texp (s)  &  Mag  & ErMag\\
\hline\hline
\endhead
\hline
\endfoot
%UUU
0.4780   & Dk1.54m+DFOSC    & U        & 9000      & 23.09 & 0.18 \\
1.5939   & 4.0mKPNO         & U        & 1600      & 23.75 & 0.26 \\
\hline
%BBB
0.1297   & 6.0mBTA+SCORPIO  & B        & 600       & 22.37 & 0.05 \\
0.1385   & 6.0mBTA+SCORPIO  & B        & 500       & 22.44 & 0.05 \\
0.1467   & 6.0mBTA+SCORPIO  & B        & 600       & 22.46 & 0.05 \\
0.1554   & 6.0mBTA+SCORPIO  & B        & 600       & 22.55 & 0.05 \\
0.2301   & 6.0mBTA+SCORPIO  & B        & 500       & 22.79 & 0.07 \\
0.3712   & Dk1.54m+DFOSC    & B        & 1200      & 23.05 & 0.10 \\
0.5681   & Dk1.54m+DFOSC    & B        & 1200      & 23.54 & 0.17 \\
1.5001   & Dk1.54m+DFOSC    & B        & 9000      & 24.26 & 0.10 \\
242.5213 & 3.5mCAHA         & B        & 1800      & 25.32\footnote{Host galaxy magnitude} & 0.22 \\
\hline
%VVV
0.2245   & 6.0mBTA+SCORPIO  & V        & 300       & 22.11 & 0.06 \\
0.3599   & Dk1.54m+DFOSC    & V        & 600       & 22.24 & 0.09 \\
1.6206   & 4.0mKPNO         & V        & 1200      & 23.98 & 0.22 \\
2.4394   & 4.0mCTIO         & V        & 1500      & 24.10 & 0.17 \\
\hline
%RRR
0.0155   & Kiso 1.05 Schmidt& Rc        & 300       &$>$19.50& ---  \\
0.0388   & 1.0mLOT          & Rc\footnote{{\it VR} broad band filter
was transformed to {\it Rc}.}        & 180       & 20.34 & 0.10 \\
0.0859   & 1.0mZeiss        & Rc        & 900       & 20.99 & 0.09 \\
0.0919   & 1.0mLOT          & Rc$^{5}$  & 180       & 20.80 & 0.18 \\
0.0956   & 1.0mLOT          & Rc$^{5}$  & 300       & 20.98 & 0.14 \\
0.1029   & 6.0mBTA+SCORPIO  & Rc        & 180       & 20.92 & 0.03 \\
0.1068   & 6.0mBTA+SCORPIO  & Rc        & 180       & 20.94 & 0.03 \\
0.1104   & 6.0mBTA+SCORPIO  & Rc        & 180       & 20.92 & 0.03 \\
0.1139   & 6.0mBTA+SCORPIO  & Rc        & 180       & 20.92 & 0.02 \\
0.1172   & 6.0mBTA+SCORPIO  & Rc        & 180       & 20.95 & 0.02 \\
0.1207   & 6.0mBTA+SCORPIO  & Rc        & 180       & 21.03 & 0.03 \\
0.1240   & 6.0mBTA+SCORPIO  & Rc        & 180       & 21.05 & 0.03 \\
0.1471   & RTT150+TFOSC     & Rc        & 450       & 21.13 & 0.08 \\
0.1629   & RTT150+TFOSC     & Rc        & 450       & 21.05 & 0.06 \\
0.1771   & RTT150+TFOSC     & Rc        & 540       & 21.10 & 0.06 \\
0.1863   & RTT150+TFOSC     & Rc        & 540       & 21.19 & 0.06 \\
0.1984   & RTT150+TFOSC     & Rc        & 480       & 21.42 & 0.08 \\
0.2080   & RTT150+TFOSC     & Rc        & 480       & 21.40 & 0.08 \\
0.2175   & RTT150+TFOSC     & Rc        & 480       & 21.39 & 0.09 \\
0.2204   & 6.0mBTA+SCORPIO  & Rc        & 180       & 21.32 & 0.04 \\
0.2296   & RTT150+TFOSC     & Rc        & 480       & 21.19 & 0.06 \\
0.2380   & RTT150+TFOSC     & Rc        & 480       & 21.52 & 0.09 \\
0.2463   & RTT150+TFOSC     & Rc        & 480       & 21.56 & 0.08 \\
0.2742   & RTT150+TFOSC     & Rc        & 720       & 21.52 & 0.06 \\
0.2867   & RTT150+TFOSC     & Rc        & 720       & 21.58 & 0.06 \\
0.3392   & RTT150+TFOSC     & Rc        & 600       & 21.80 & 0.12 \\
0.3496   & RTT150+TFOSC     & Rc        & 600       & 21.50 & 0.09 \\
0.3540   & Dk1.54m+DFOSC    & Rc        & 300       & 21.81 & 0.13 \\
0.3600   & RTT150+TFOSC     & Rc        & 600       & 21.84 & 0.14 \\
0.3705   & RTT150+TFOSC     & Rc        & 600       & 21.95 & 0.15 \\
0.4362   & Dk1.54m+DFOSC    & Rc        & 600       & 21.91 & 0.08 \\
0.5366   & Dk1.54m+DFOSC    & Rc        & 900       & 22.00 & 0.07 \\
0.5963   & Dk1.54m+DFOSC    & Rc        & 1200      & 22.02 & 0.08 \\
0.7166   & 2.0mFTN          & Rc        & 2400      & 22.50 & 0.11 \\
1.4024   & Dk1.54m+DFOSC    & Rc        & 5700      & 23.08 & 0.09 \\
1.5864   & Dk1.54m+DFOSC    & Rc        & 3600      & 23.31 & 0.14 \\
2.4911   & Dk1.54m+DFOSC    & Rc        & 14700     & 23.86 & 0.17 \\
3.4506   & Dk1.54m+DFOSC    & Rc        & 14000     & 23.85 & 0.13 \\
5.1575   & RTT150+ANDOR CCD & Rc        & 9000      & 23.70 & 0.20 \\
5.5124   & Dk1.54m+DFOSC    & Rc        & 7200      & 23.72 & 0.18 \\
7.4663   & Dk1.54m+DFOSC    & Rc        & 15600     & 23.97 & 0.11 \\
30.2567  & 3.5mTNG          & Rc        & 3600      & 24.64 & 0.17 \\
242.4885 & 3.5mCAHA         & Rc        & 2500      & 24.60$^4$ & 0.15 \\
\hline
%III
0.3871   & Dk1.54m+DFOSC    & Ic        & 1200      & 21.25 & 0.07 \\
0.6137   & Dk1.54m+DFOSC    & Ic        & 1500      & 22.52 & 0.11 \\
6.4766   & Dk1.54m+DFOSC    & Ic        & 13800     & 23.50 & 0.18 \\
242.5446 & 3.5mCAHA         & Ic        & 1500      &$>$24.0& ---  \\
\hline
%Unfiltered
-0.00340 & WIDGET           & Unfiltered& 95        &$>$9.7& ---  \\
-0.00224 & WIDGET           & Unfiltered& 95        &$>$9.7& ---  \\
-0.00108 & WIDGET           & Unfiltered& 95        &$>$9.7& ---  \\
0.00007  & WIDGET           & Unfiltered& 95        &$>$9.8& ---  \\
0.00123  & WIDGET           & Unfiltered& 95        &$>$9.7& ---  \\
\end{longtable}
}

\end{appendix}
\end{document}